\documentclass[sn-mathphys-num]{sn-jnl}
\usepackage[utf8]{inputenc}

\usepackage{graphicx} 

\usepackage{hyperref}
\hypersetup{
  colorlinks,%
  citecolor=black,%
  filecolor=black,%
  linkcolor=black,%
  urlcolor=blue
  }
  
\usepackage{amssymb}
\usepackage{amsmath}

\usepackage{subfigure}
\usepackage{wrapfig}  

\usepackage{soul}
\usepackage[mathlines]{lineno}

\newcommand{\emult}{{$\left<\dd N_{\mathrm{ch}}/\dd \eta\right>$}}%
\newcommand{\emultIncl}{{$\left<\dd N_{\mathrm{ch,Incl}}/\dd \eta\right>$}}%

\newcommand{\mpT}{{$\left<p_{\rm T}\right>$} }%

\newcommand{\pt}{p_{\rm T}}
\newcommand{\pp}{\textit{pp} }
\newcommand{\dd}{ {\mathrm d} }
\usepackage{orcidlink}
\newcommand{\orcidA}{\orcidlink{0000-0003-2513-2459}} 
\newcommand{\orcidB}{\orcidlink{0000-0003-2849-0120}} 
\newcommand{\orcidC}{\orcidlink{0000-0001-9223-6480}} 
\newcommand{\orcidD}{\orcidlink{0000-0003-4749-5250}} 

\begin{document}

\title{Investigating the soft and hard limits in transverse momentum spectra in \textit{pp} collisions.} 

\author*[1,2]{\fnm{G\'abor} \sur{B\'ir\'o}\orcidB{}}
\email{biro.gabor@wigner.hun-ren.hu}
\equalcont{These authors contributed equally to this work as co-first authors.}

\author[3]{\fnm{Leonid} \sur{Serkin}\orcidD{}}
\email{lserkin@ciencias.unam.mx} 
\equalcont{These authors contributed equally to this work as co-first authors.}

\author[4]{\fnm{Guy} \sur{Pai\'c}\orcidA{}}
\email{Guy.Paic@cern.ch} 

\author[1]{\fnm{Gergely G\'abor} \sur{Barnaf\"oldi}\orcidC{}}
\email{barnafoldi.gergely@wigner.hun-ren.hu}

\affil[1]{HUN-REN Wigner Research Centre for Physics, 29--33 Konkoly--Thege Mikl\'os Rd., H-1121 Budapest, Hungary}
\affil[2]{ELTE Eötvös Loránd University, Institute of Physics, Pázmány Péter Sétány 1/A, H-1117  Budapest, Hungary}
\affil[3]{Facultad de Ciencias, Universidad Nacional Autónoma de México, Circuito Exterior s/n, Ciudad Universitaria, Coyoacán, Ciudad de México, 04510, México}
\affil[4]{Instituto de Ciencias Nucleares, Universidad Nacional Autónoma de México,  Apartado Postal 70-543, Ciudad de México 04510, México}

\date{\today}

\abstract{
The transverse momentum spectra and their multiplicity dependence serve as key tools for extracting parameters to be compared with theoretical models. Over the past decade, the scientific community has extensively studied the possibility of a system analogous to quark-gluon plasma, predicted in heavy nuclei collisions, also existing in collisions involving light nuclei and protons. 
We have reanalysed the data published by the ALICE Collaboration at the LHC. We present the dependence of the mean transverse momenta obtained in the soft and soft+hard (mixed) parts. Finally, we also discuss possible refinements of the analyses concerning the use of statistical parameters of higher order, aimed at a more detailed way of comparing the models with data.
}


\maketitle

\section{Introduction}
\label{sec:introduction}
The investigation of collective effects in proton-proton (\textit{pp}) collisions has been a highly active area of research over the past decade, with numerous studies arguing the presence of collective flow in \pp collisions~\cite{ALICE:2016fzo, CMS:2016fnw, ATLAS:2018ngv, ALICE:2022wpn, Nagle:2018nvi,  STRICKLAND201992, Mishra:2022kre, Bierlich:2022ned}. Recent measurements of high-multiplicity events in both \pp and proton-nucleus collisions have raised new questions regarding the potential production of hot QCD matter in small systems, see e.g. \cite{Bozek:2013uha, osti_1188210, Ghosh:2014eqa, Gu:2022xjn, Ortiz:2016kpz, BLOK2019259,PhysRevC.101.064902, Ji:2023eqn} and the references therein. This is particularly intriguing given that in high-multiplicity \pp collision events, the energy density may reach levels comparable to those observed in nucleus-nucleus collisions. Hence, it is essential to investigate both traditional and novel observables across various event multiplicity classes~\cite{Ortiz:2022zqr, Ortiz:2022mfv, Biro:2020kve, Mishra:2021hnr, Mishra:2018pio, Horvath:2023lho}.

Inspired by the results obtained by the ALICE Collaboration~\cite{ALICE:2010syw, ALICE:2010cin, ALICE:2013txf, ALICE:2015qqj,  ALICE:2018pal, ALICE:2018vuu, ALICE:2019dfi, ALICE:2020jsh, ALICE:2020nkc}, we notice that, traditionally, collision data observables are presented as \textit{mean} values of more differential distributions, such as mean charged-particle multiplicity $\left<  N_{\mathrm{ch}} \right>$, mean of transverse momentum $\left<p_{\rm T}\right>$, mean values of anisotropic flow, etc. 
However, an accurate interpretation of theoretical model parameters from measured data may encounter inconsistencies due to the QCD theory's dual, ``Janus-faced'' nature, embracing both perturbative and nonperturbative, or soft and hard, aspects. 

The comparison of the mean values in data and model predictions gives at best ``satisfactory'' agreement for some models. However, the agreement or disagreement of the \textit{means} of a model with the data may not really point to the details of the result, given that different models may have the same means as the data but with different underlying assumptions. An important feature of the means is the range used for averaging. In the case of exponentially decreasing spectra the information is mainly influenced by the lower $\pt$ part, while the features of the highest part are not playing an important role. In the present paper we have tried to investigate the lowest $\pt$ dominated by the soft part of the interaction, extensively studied as the main source of collective effects in \pp collisions~\cite{Gu:2022xjn, Ortiz:2016kpz, Liu:2022ikt,Bierlich:2017vhg,Zhao:2017rgg}.


\section[Experimental {\it vs}. Monte Carlo pT spectra]{Experimental {\it vs}. Monte Carlo \( p_{\text{T}} \) spectra}
\label{sec:expdat}

\begin{figure}[ht]
\centering
\includegraphics[width=0.49\linewidth]{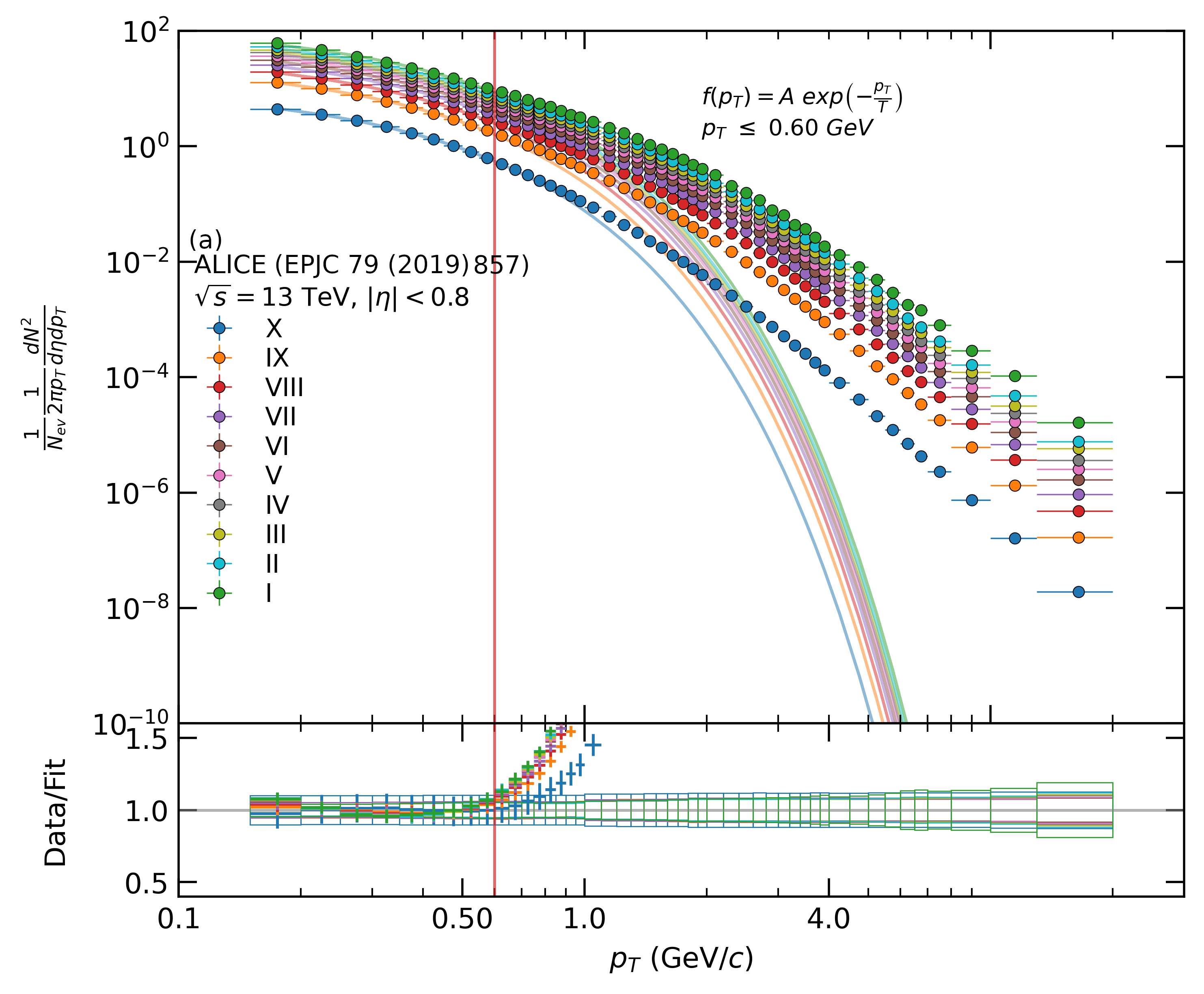}
\includegraphics[width=0.49\linewidth]{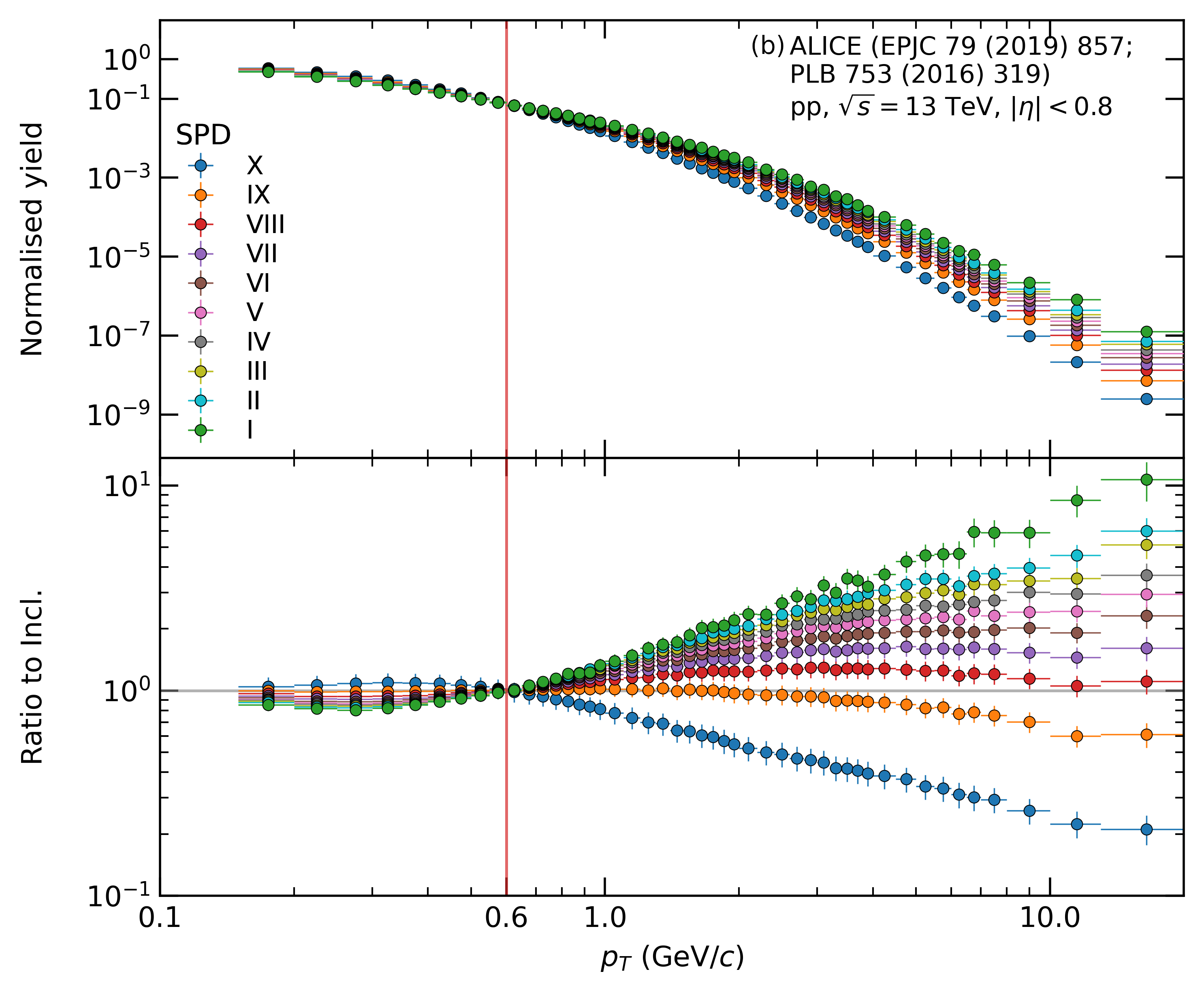}
\caption{
Simple exponential fits (left) and normalized transverse momentum ratios with respect to the inclusive spectrum (right) as a function of the charged-particle pseudorapidity density as measured by the ALICE Collaboration in inelastic \pp collisions at $\sqrt{s}$~=~ 13 TeV at various event multiplicity classes (X is the lowest, I is the highest---for the specific values see~\cite{ALICE:2015qqj, ALICE:2019dfi}). 
}
\label{fig:pT_spectra13}
\end{figure}

\begin{figure*}[h]
  \centering
  \includegraphics[width=0.48\linewidth]{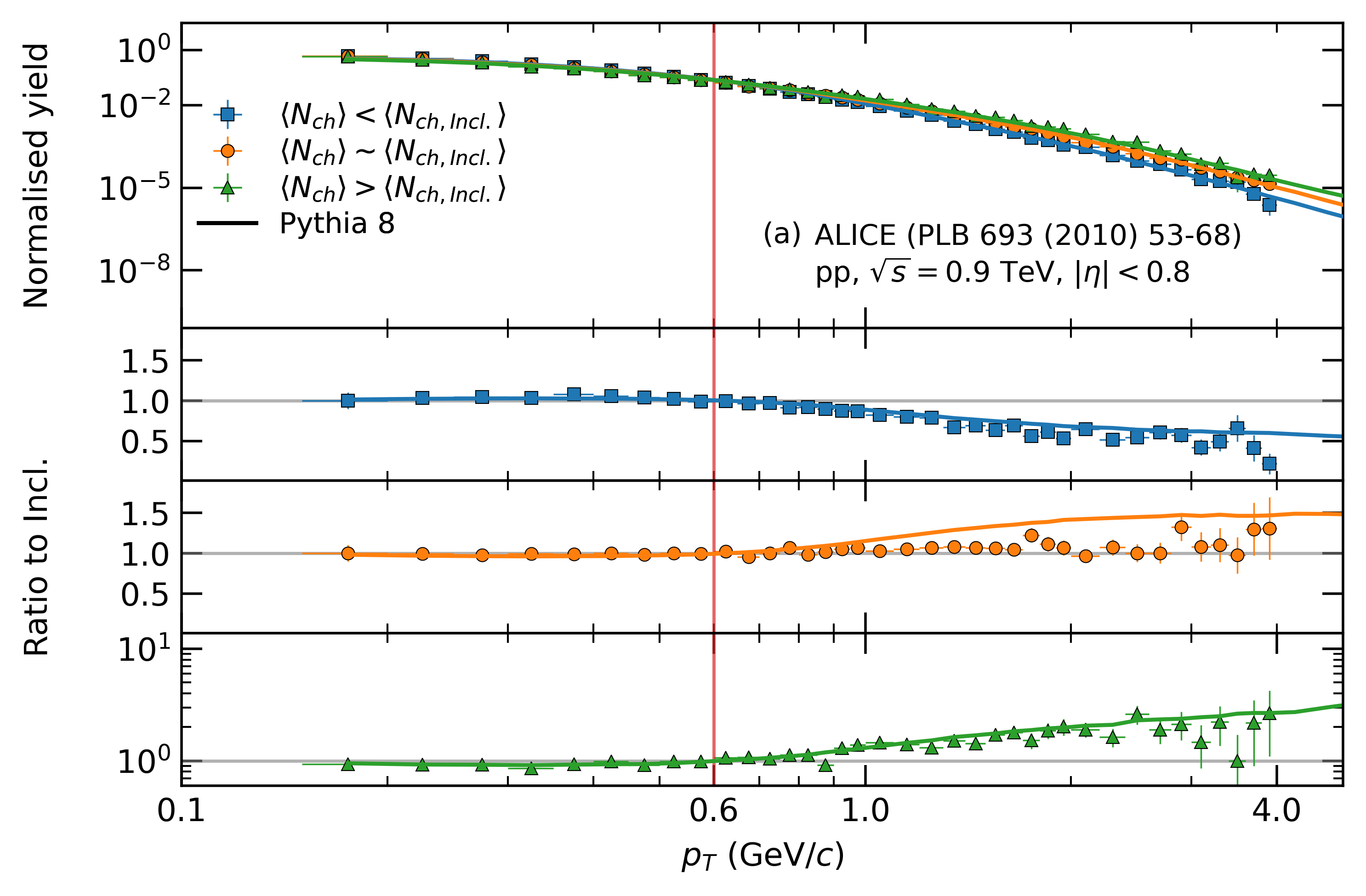}
  \includegraphics[width=0.48\linewidth]{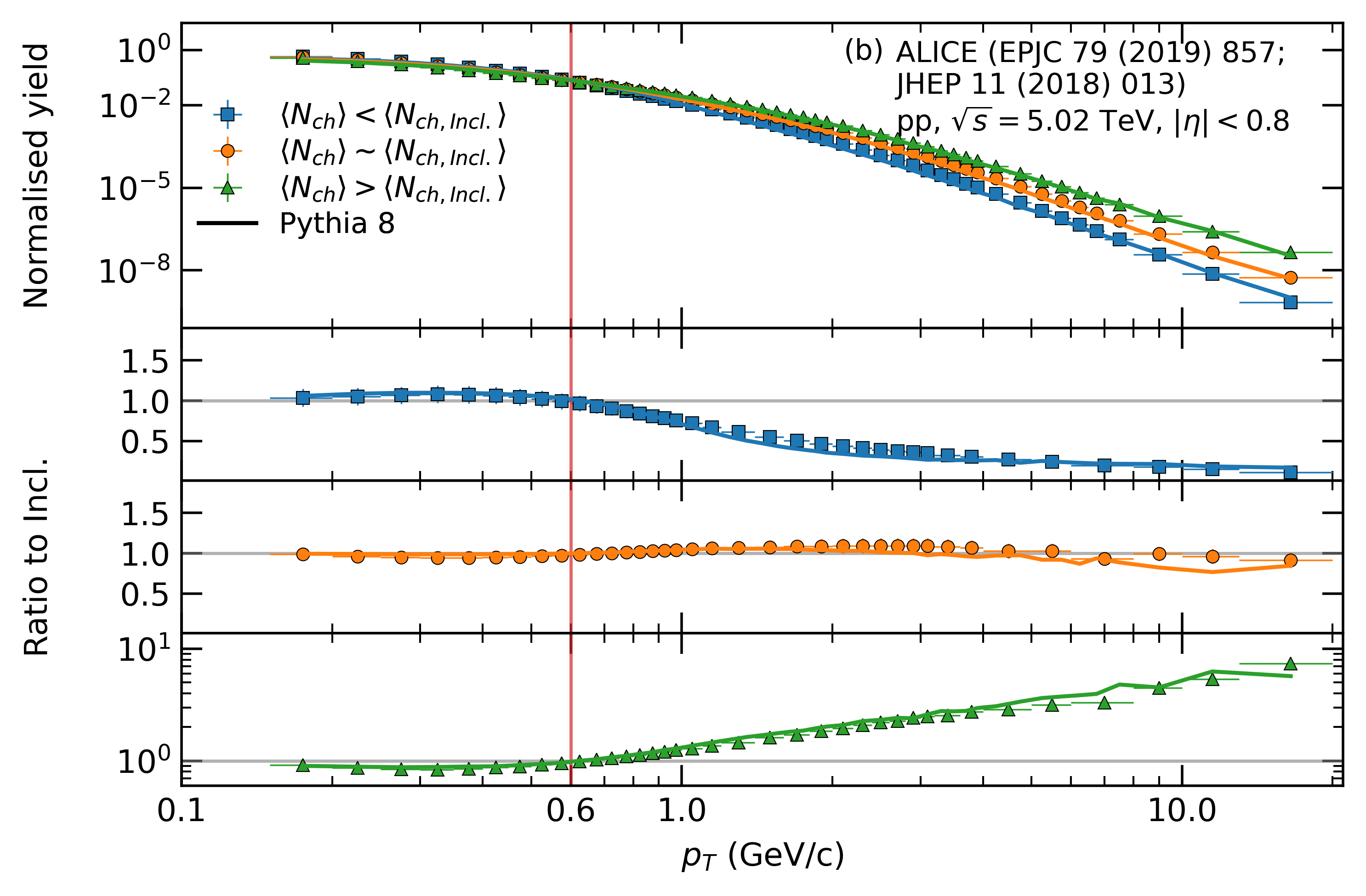}
  \includegraphics[width=0.48\linewidth]{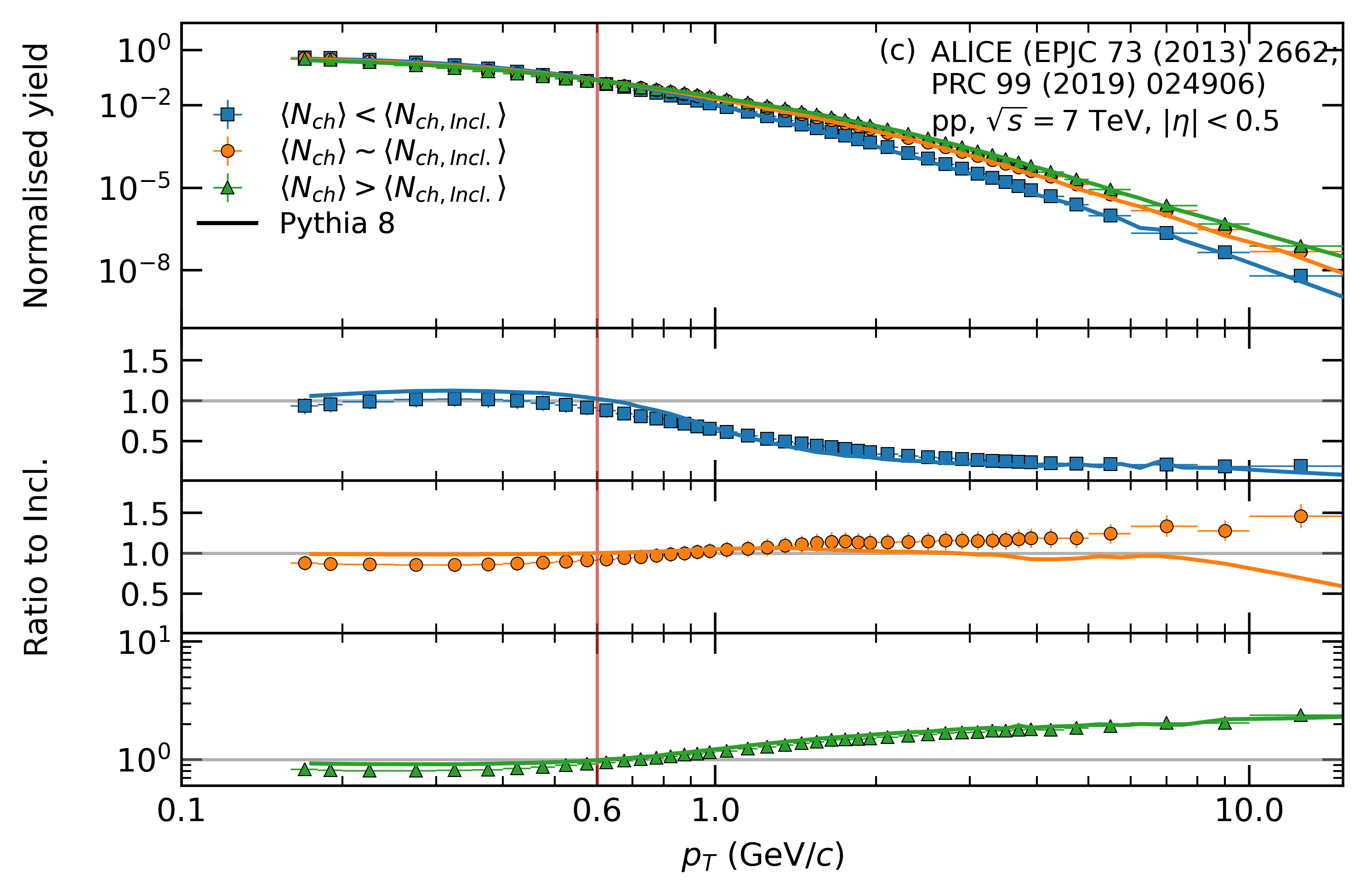}
  \includegraphics[width=0.48\linewidth]{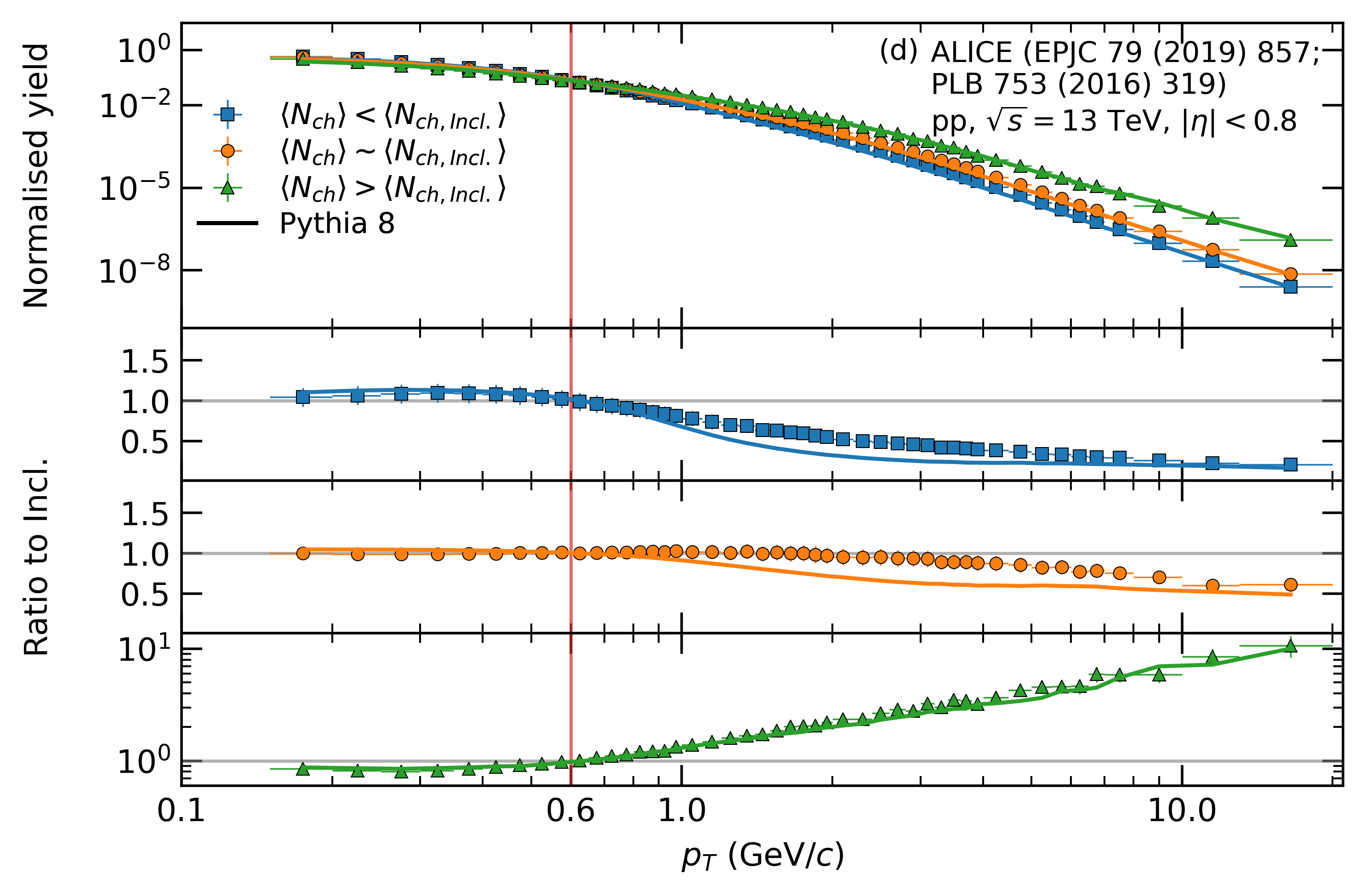}
  \caption{
$N_{\mathrm{ch}}$-integrated transverse momentum distributions of charged particles for three different multiplicity classes as measured by the ALICE Collaboration~\cite{ALICE:2010syw, ALICE:2010cin, ALICE:2013txf, ALICE:2015qqj,  ALICE:2018pal, ALICE:2018vuu, ALICE:2019dfi, ALICE:2020jsh, ALICE:2020nkc}. Results for \pp collisions at (a) $\sqrt{s} = 0.9$ TeV, (b) $\sqrt{s} = 5.02$ TeV, (c) $\sqrt{s} = 7$ TeV and (d) $\sqrt{s} = 13$ TeV are presented. The top panels display ratios of the $\pt$ spectra to the inclusive spectrum for the lowest-multiplicity class. Similarly, the middle panels show ratios of the $\pt$ spectra to the inclusive spectrum for the multiplicity class closest to the inclusive mean value. In the bottom panels, ratios of the $\pt$ spectra to the inclusive spectrum are displayed on a logarithmic scale for the highest-multiplicity class. The solid line in each figure represents the {\sc Pythia~8}  simulation at the corresponding $\sqrt{s}$. 
}
\label{fig:pp_normed2}
\end{figure*}

In recent years, the ALICE Collaboration has measured the charged-particle transverse momentum spectra as a function of charged-particle multiplicity at various centre-of-mass energies, $\sqrt{s}$. A parametrization over the full-$\pt$ range can be achieved well with the Tsallis function~\cite{Tsallis:1987eu} or the modified Hagedorn function~\cite{Hagedorn:1983wk}. Our particular interest lies in exploring the soft, low-$\pt$ region of the spectrum, known to have an approximately exponential shape~\cite{Garcia:2022eqg}. 

Recent studies showed that traditional, Boltzmann\,--\,Gibbs statistics-based effective models are able to take into account the enhancement of soft pions, and therefore give a good description of the experimental data up to 4-5 GeV/$c$~\cite{Blaschke:2020afk}. However, exponential functions are not able to fit the high-$\pt$ data by nature. The left panel of Fig.~\ref{fig:pT_spectra13}, presenting the multiplicity dependent charged hadron transverse momentum in $\sqrt{s}=13$ TeV (where X denotes the lowest, and I the highest multiplicity class), illustrates that a Boltzmann function results in a satisfactory fit in the low $\pt$ spectrum. Beyond $\pt \sim$ few GeV/$c$, the goodness of the fits break down rapidly. Supposing that the region where a Boltzmannian fit gives a satisfactory fit correspond to soft interactions, one sees that the spectra may be roughly separated into three regions: the low $\pt$ part representing purely soft interaction, an intermediate part where the hard interactions get mixed with soft ones, and finally beyond around 4 GeV/$c$ a region where we have contribution exclusively from hard interactions. Therefore, the usual observable quantities that are based on cut-distributions must be interpreted with caution. In the right panel yields in multiplicity classes are presented relative to the inclusive yield. In comparison to the left panel, an interesting observation is: both the best fit for the low-$\pt$ range and the centrality-class normalized yields select a natural  value at $\pt \approx 0.6$~GeV/$c$. This observation motivated us further to investigate the two regions  (before and after the crossing points) separately.

The top panels of Fig.~\ref{fig:pp_normed2} present the $N_{\mathrm{ch}}$-integrated $\pt$ distributions of unidentified charged particles at mid-pseudorapidity ($|\eta| < 0.5$ or $|\eta| < 0.8$), measured by the ALICE detector in inelastic (INEL) \pp collisions at $\sqrt{s}$~=~0.9, 5.02, 7 and 13 TeV. For better visibility, in this case only 3 multiplicity classes are shown, as in the right panel of Fig.~\ref{fig:pT_spectra13}. Data by the ALICE Collaboration were used~\cite{ALICE:2010syw, ALICE:2010cin, ALICE:2013txf, ALICE:2015qqj,  ALICE:2018pal, ALICE:2018vuu, ALICE:2019dfi, ALICE:2020jsh, ALICE:2020nkc}, where charged-particle pseudorapidity densities, \emult, are based on tracklets in the Silicon Pixel detector, and multiplicities recorded in the forward V0 detectors (V0M amplitude). The values of the \emultIncl~in the inclusive ($\mathrm{INEL} > 0$) multiplicity class are:
$4.20 \pm 0.03$ at $\sqrt{s} = 0.9$~TeV and $|\eta| < 0.5$~\cite{ALICE:2010cin}, 
$5.91 \pm 0.45$ at $\sqrt{s} = 5.02$~TeV and $|\eta| < 0.8$~\cite{ALICE:2019dfi}, 
$5.96 \pm 0.23$ at $\sqrt{s} = 7$~TeV and $|\eta| < 0.5$~\cite{ALICE:2018pal}, 
and $7.60 \pm 0.50$ at $\sqrt{s} = 13$~TeV and $|\eta| < 0.8$~\cite{ALICE:2019dfi}, respectively.

For each collision energy, several multiplicity classes are considered, and then for sake of clarity  we report only  spectra for three multiplicity region: spectra corresponding to values lower (\emult$<$\emultIncl), approximately equal (\emult$\sim$\emultIncl), and higher (\emult$>$\emultIncl) than the mean value of the multiplicity for the inclusive case.

First, we compare the experimental results with {\sc Pythia~8} (version 8.309) Monte Carlo (MC) simulation~\cite{Sjostrand:2014zea}, widely used for studying LHC physics. We generated $\sim 50$~million inelastic events at each $\sqrt{s}$, including non-diffractive and diffractive components, with the latter represent 31\% and 27\% of the total cross-section at $\sqrt{s}$~=~0.9 and 13 TeV, respectively. We employed the Monash 2013 tune, calibrated using early LHC measurements~\cite{Skands:2014pea}, and processed the MC simulation with {\sc Rivet v3}~\cite{Bierlich:2019rhm}. Simulations were also carried out employing a core-corona approach based on new microcanonical hadronisation procedures in the recently released {\sc Epos~4} (version 4.0.0) framework~\cite{Werner:2023zvo, Werner:2023fne}. {\sc Epos~4} incorporates collective (flow-like) effects and integrates knowledge from $S$-matrix theory, perturbative QCD and saturation. Following the definition provided in \cite{ALICE:2019dfi}, the results are presented for primary charged particles in the kinematic range $0.15$~GeV/$c$ $\leq \pt \leq 20$~GeV/$c$. As can be seen in Fig.~\ref{fig:pp_normed2}, {\sc Pythia~8} predictions successfully describe the qualitative features of the evolution of the $\pt$ spectra, in particular at low values.

Following Ref.~\cite{ALICE:2015qqj}, the bottom panels of Fig.~\ref{fig:pp_normed2} present the ratios of the $\pt$ spectra to the inclusive spectrum for the three multiplicity classes. In each case, the spectra were normalised by the integral prior to dividing. The features of the $\pt$ spectra, namely the change in spectral shape from low- to high-multiplicity values, are qualitatively consistent across all energies. The only significant difference is the higher multiplicity reach at $\sqrt{s}$~=13 TeV compared to that at 0.9 TeV. 
By normalizing the spectra, the analysis isolates differences in the shape of the $\pt$ spectra, eliminating the influence of absolute particle yields. This is crucial for understanding the underlying physics affecting particle production dynamics as a function of $\pt$. Surprisingly, the behaviour of the ratios to the inclusive spectrum for the three multiplicity classes resulted in a singular pattern, exhibiting a crossing at the unity value at $\sim$0.6~GeV/$c$ at all studied $\sqrt{s}$~energies. Below this value, the ratios are almost flat and center-of-mass energy independent in the range of collision energies reported. Above this $\pt$ value, the ratios diverge significantly at all energies.

\begin{wrapfigure}{r}{0.5\textwidth}
\centering
\includegraphics[width=\linewidth]{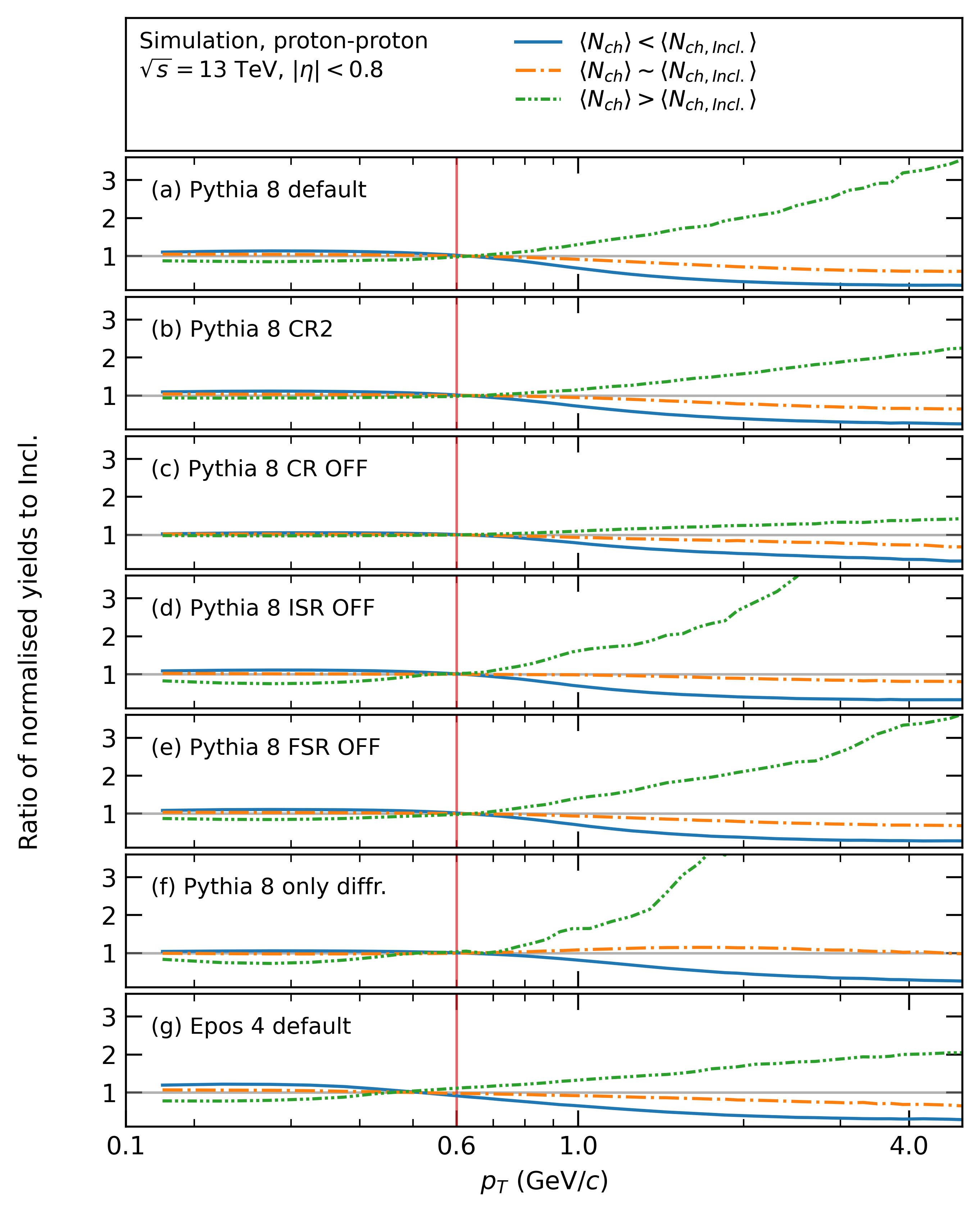}
\caption{
Ratios of the normalised $\pt$ spectra to the inclusive spectrum for the three multiplicity classes at $\sqrt{s}=13$~TeV.
Predictions from {\sc Pythia~8} simulation with different settings are shown: (a) default {\sc Pythia~8} model, (b) {\sc Pythia~8} with modified CR, (c) {\sc Pythia~8} with CR turned off, (d) {\sc Pythia~8} without ISR, (e) {\sc Pythia~8} without FSR, (f) {\sc Pythia~8} events produced only via diffractive processes. The prediction from {\sc Epos 4} is shown in (g). The red line denotes the observed crossing value at $\sim$0.6~GeV/$c$.
}
\label{fig:MC_comparisons}
\end{wrapfigure}

Different configurations within the {\sc Pythia~8} MC model were tested. Events were simulated with the default parametrisation of the so-called multi-parton interactions-based model of colour reconnection (CR)~\cite{PhysRevD.36.2019, Sjostrand:2004pf}, as well as with modified CR based on the gluon-move model~\cite{Argyropoulos:2014zoa}, without any colour reconnection, and without initial-state (ISR) or final-state radiation (FSR). Additionally, given the sensitivity of the low-$\pt$ region to diffractive contributions~\cite{CMS:2020dns}, events produced only via diffractive processes were also tested. 

Figure~\ref{fig:MC_comparisons} presents the ratios of the normalised $\pt$ spectra to the inclusive spectrum for the three multiplicity classes at $\sqrt{s}=13$~TeV using different {\sc Pythia~8} settings, along with the prediction from {\sc Epos 4}. The ratios present the same behaviour as it has been observed earlier: there is a significant hardening of the $\pt$ distributions towards larger multiplicities, resulting in a change of the spectral shape compared to the inclusive spectrum and attributed to the presence of energetic jets~\cite{ALICE:2010syw, ALICE:2013rdo, ALICE:2015qqj}. 
However, the crossing point is practically independent of the different {\sc Pythia~8} settings. A deviation in terms of the crossing point at the unity value is observed only for the {\sc Epos~4} case.

It is well known that transverse momentum distributions for different multiplicities result from distinct contributions: the low-$\pt$ part of the spectrum is primarily represented by soft contributions and diffractive processes, whereas the high-$\pt$ part is attributed to a mixture of jet and minijet contributions~\cite{ALICE:2022xip, ALICE:2023plt, Ortiz:2023slm}. The boundary between soft and hard processes is not precisely defined but is typically assumed to fall within the range of $1-2$~GeV/$c$~\cite{Prasad:2015uma, PhysRevC.64.034901, ATLAS:2016zba,Garcia:2022eqg}.

Our observation of a crossing-point, which is remarkably close to the breakdown of the naive exponential fit, drives us to the same conclusion: the $\pt$ spectra in a given multiplicity class stems from at least two different processes. Hence, it seems legitimate to separate each individual $\pt$ spectrum into two parts at the observed crossing value at $\sim$0.6~GeV/$c$:
\begin{enumerate}[label={\arabic{enumi}}]
    \item Predominantly soft region (low-$p_{\rm T}$):  $0.15$~GeV/$c$ $\leq \pt \leq 0.6$~GeV/$c$,
    \item Mixed (soft and hard) region (higher-$p_{\rm T}$): $0.6$~GeV/$c$ $< \pt \leq 4$~GeV/$c$.
\end{enumerate}
Within the limited event multiplicity range provided by A\-LI\-CE data at the four $\sqrt{s}$ energies investigated, the maximum $\pt$ range studied is capped at $4.0$ GeV/$c$. This limit is set by the highest $\pt$ bin available at the collision energy of $\sqrt{s}=0.9$ TeV~\cite{ALICE:2010syw}.

\section[Moments of the pT spectra in different regions]{Moments of the \( p_{\text{T}} \) spectra in different regions}
\label{sec:moments}

Based on our observation that the ratios of various experimental spectra across a wide range of collision energies exhibit a consistent crossing-point at approximately the same $p_{\rm T} \sim 0.6$~GeV/$c$ for all investigated event multiplicities, a pattern also observed in both {\sc Pythia~8} and {\sc Epos~4} predictions, it is imperative to conduct a comprehensive investigation into the statistical properties of these two distinct $p_{\rm T}$ regions: the region below the crossing-point and the region above it.

In the following, we explore the first two statistical moments of the $p_{\rm T}$ spectra, i.e. their mean and variance, as a function of the charged-particle pseudorapidity density measured by the \mbox{ALICE} Collaboration. The mean transverse momentum, \mpT\, a widely used metric in scientific literature, is defined by Eq.~\eqref{eq:mean} and calculated using a weighted averaging approach, which involves computing the weighted average of the values present in the $p_{\rm T}$ histogram. Variance, $\sigma^2$, defined in Eq.~\eqref{eq:var} as the second central moment of the $p_{\rm T}$ distribution, quantifies the variability within the data and offers insights into the distribution's shape and deviation from the mean:
\begin{eqnarray}
  \left< p_{T}\right> &=& \sum w p \left/ \sum w     \right. ,         \label{eq:mean}   \\
  \sigma^2 &=& \frac{\sum(wp^2) \sum(w) - \sum(wp)^2 }{\sum(w)^2 - \sum(w^2)}, \label{eq:var} 
\end{eqnarray}
where $p$ and $w$ represent the values of momentum at the centre of the bin, and the bin weights of the $p_{\rm T}$ histogram, respectively. The sums are running over all particles. Both the mean and variance of the $\pt$ spectra are evaluated within three regions: one below the crossing-point, one above it, and within the full-$\pt$ region studied ($0.15$~GeV/$c$ $\leq \pt \leq 4.0$~GeV/$c$). 

\begin{figure}[ht]
\centering
\includegraphics[height=0.40\linewidth]{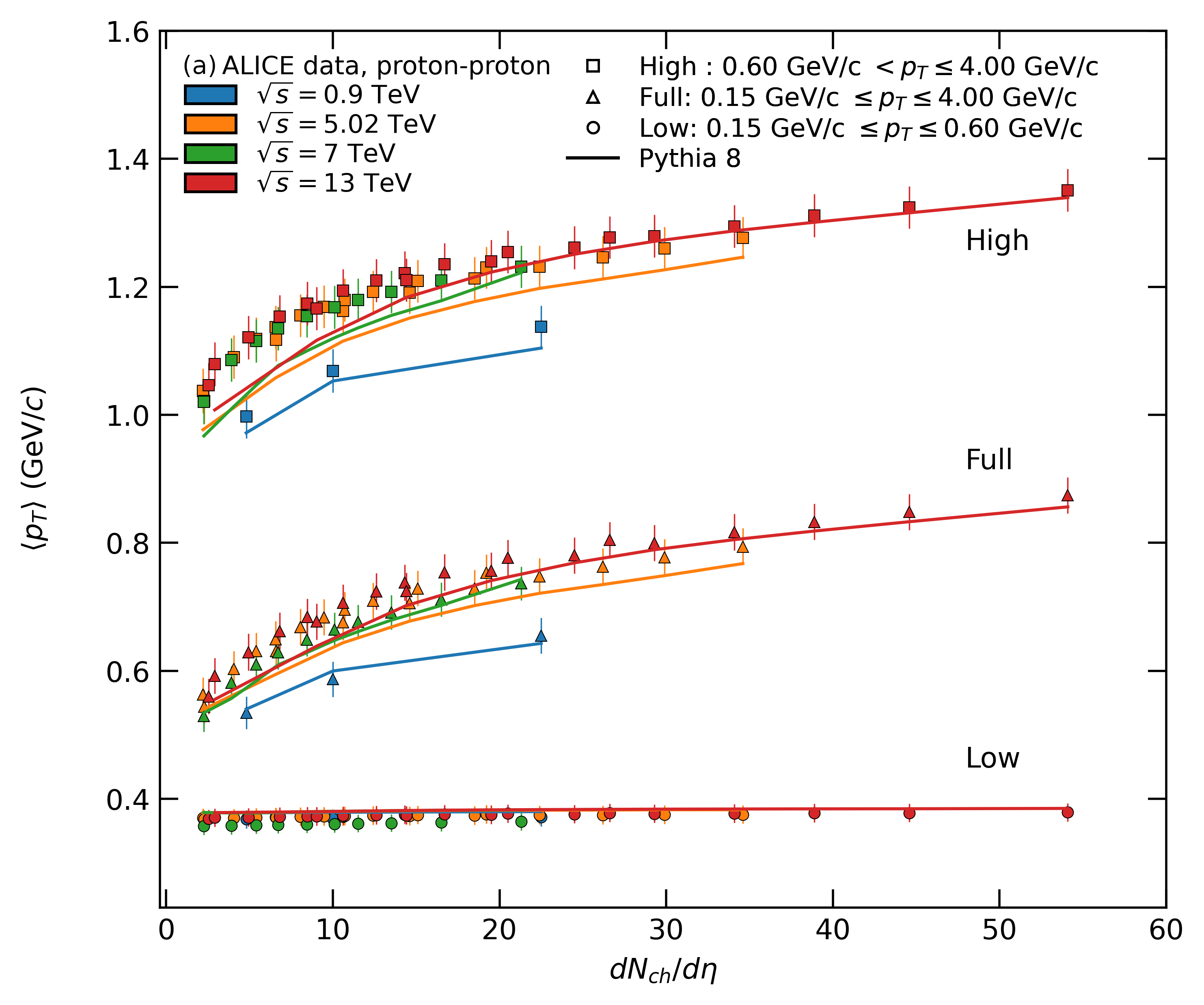}
\includegraphics[height=0.40\linewidth]{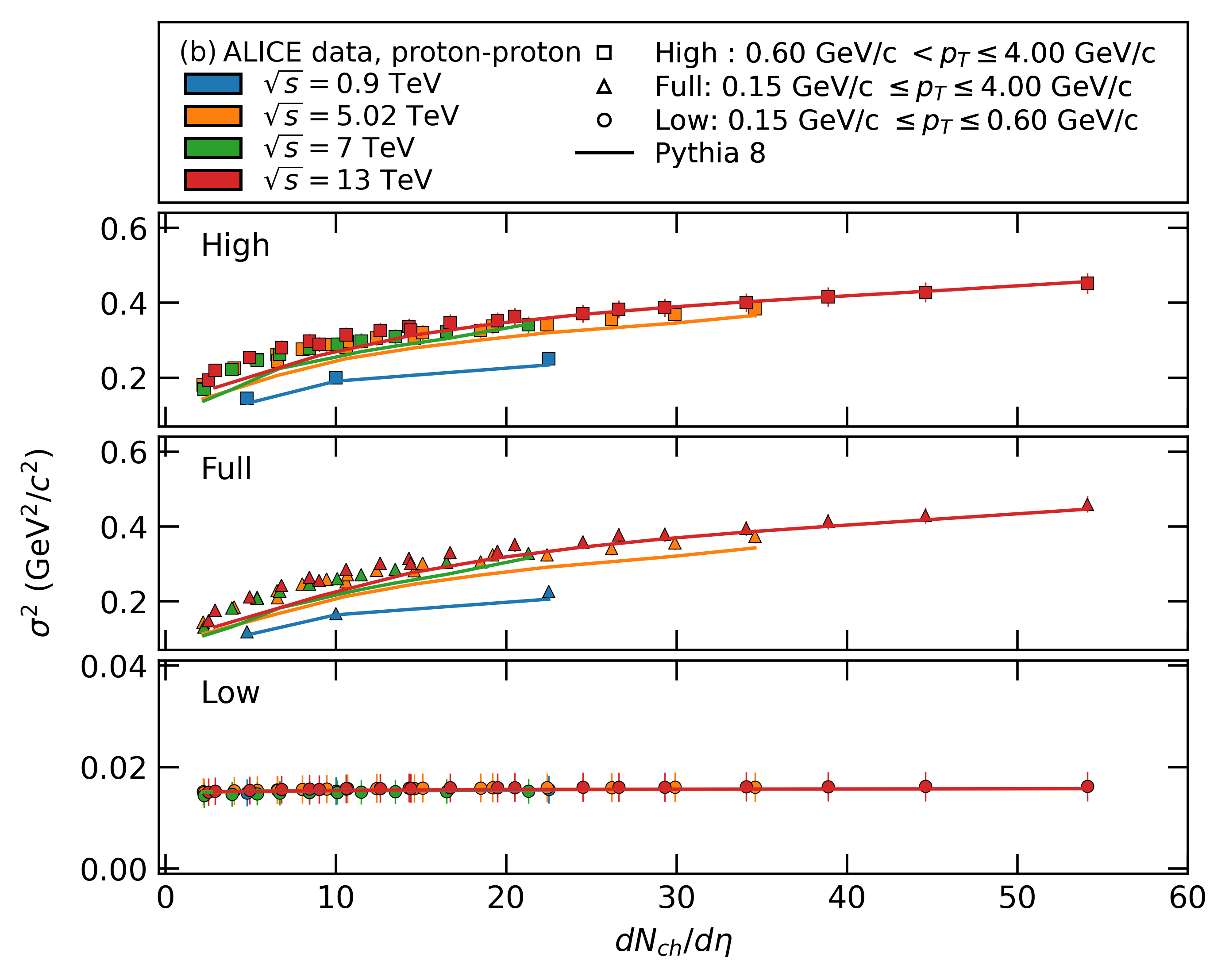}
\caption{
Mean transverse momentum (left) and variance (right) as a function of the charged-particle pseudorapidity density as measured by the ALICE Collaboration in inelastic \pp collisions at $\sqrt{s}$~=~0.9, 5.02, 7 and 13 TeV. Values of \mpT are presented for the high-$\pt$ region with $0.6$~GeV/$c$ $< \pt \leq 4$~GeV/$c$ (top panel), full-$\pt$ region with $0.15$~GeV/$c$ $\leq \pt \leq 4.0$~GeV/$c$ (middle panel) and low-$\pt$ region of $0.15$~GeV/$c$ $\leq \pt \leq 0.6$~GeV/$c$ (bottom panel). Statistical uncertainties on data are represented by error bars. The solid line in each panel represents the {\sc Pythia 8}  simulation at the corresponding $\sqrt{s}$. 
}
\label{fig:pT_means}
\end{figure}

Figure~\ref{fig:pT_means} presents the mean transverse momentum as a function of the charged-particle pseudorapidity density, as measured by the {ALICE} Collaboration in \pp collisions at $\sqrt{s}$~=~0.9, 5.02, 7 and 13 TeV. The measured values of \mpT are shown for three regions (mixed/higher-$\pt$, full-$\pt$ and soft/low-$\pt$) and compared to {\sc Pythia~8} simulations. In a similar manner, the right panel of Fig.~\ref{fig:pT_means} presents the variance of the $\pt$ spectra measured by the ALICE Collaboration at four collision energies and its comparison with MC simulation.

\begin{figure}[tb]
\centering
\includegraphics[width=0.49\linewidth]{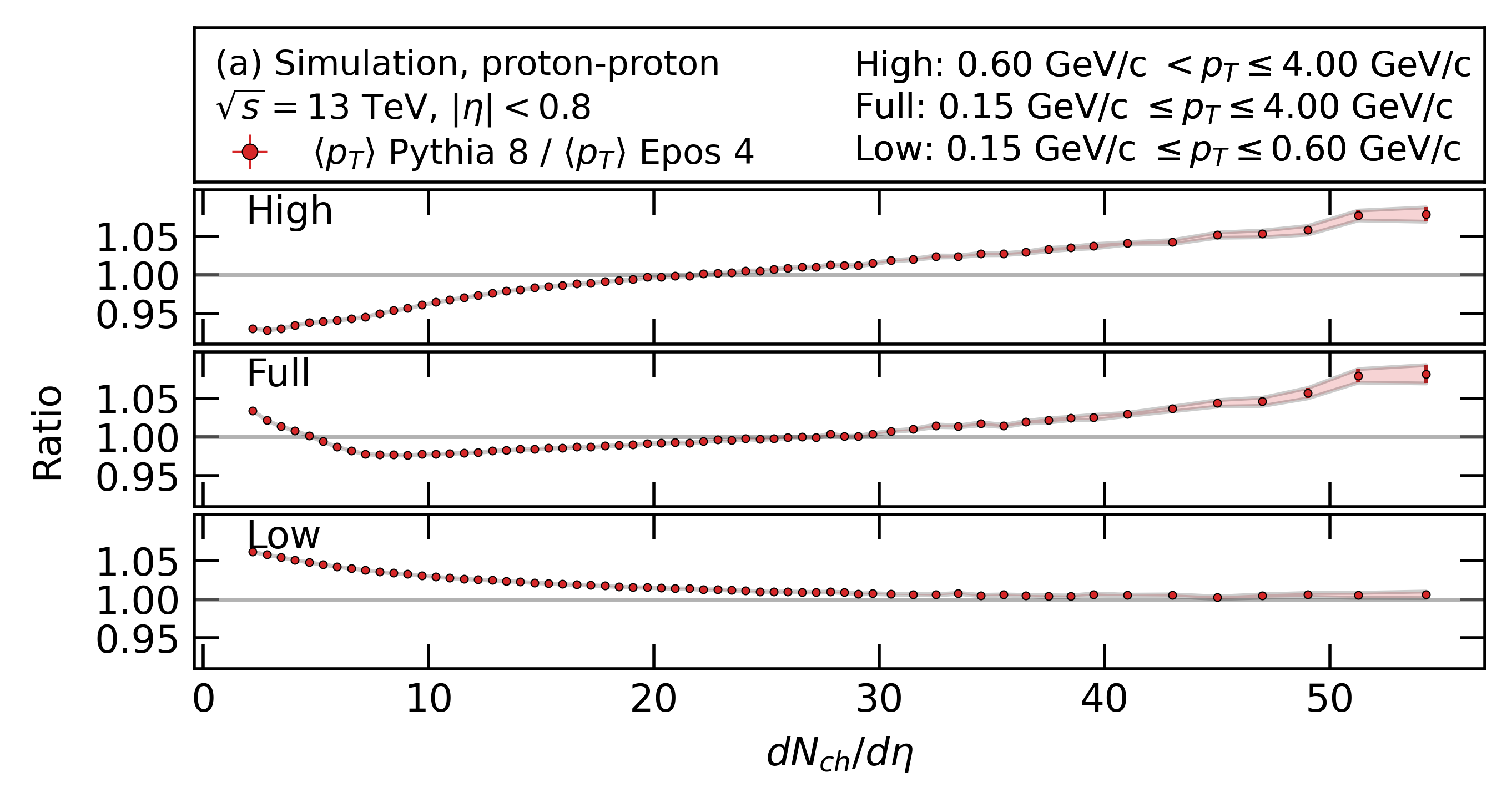}
\includegraphics[width=0.49\linewidth]{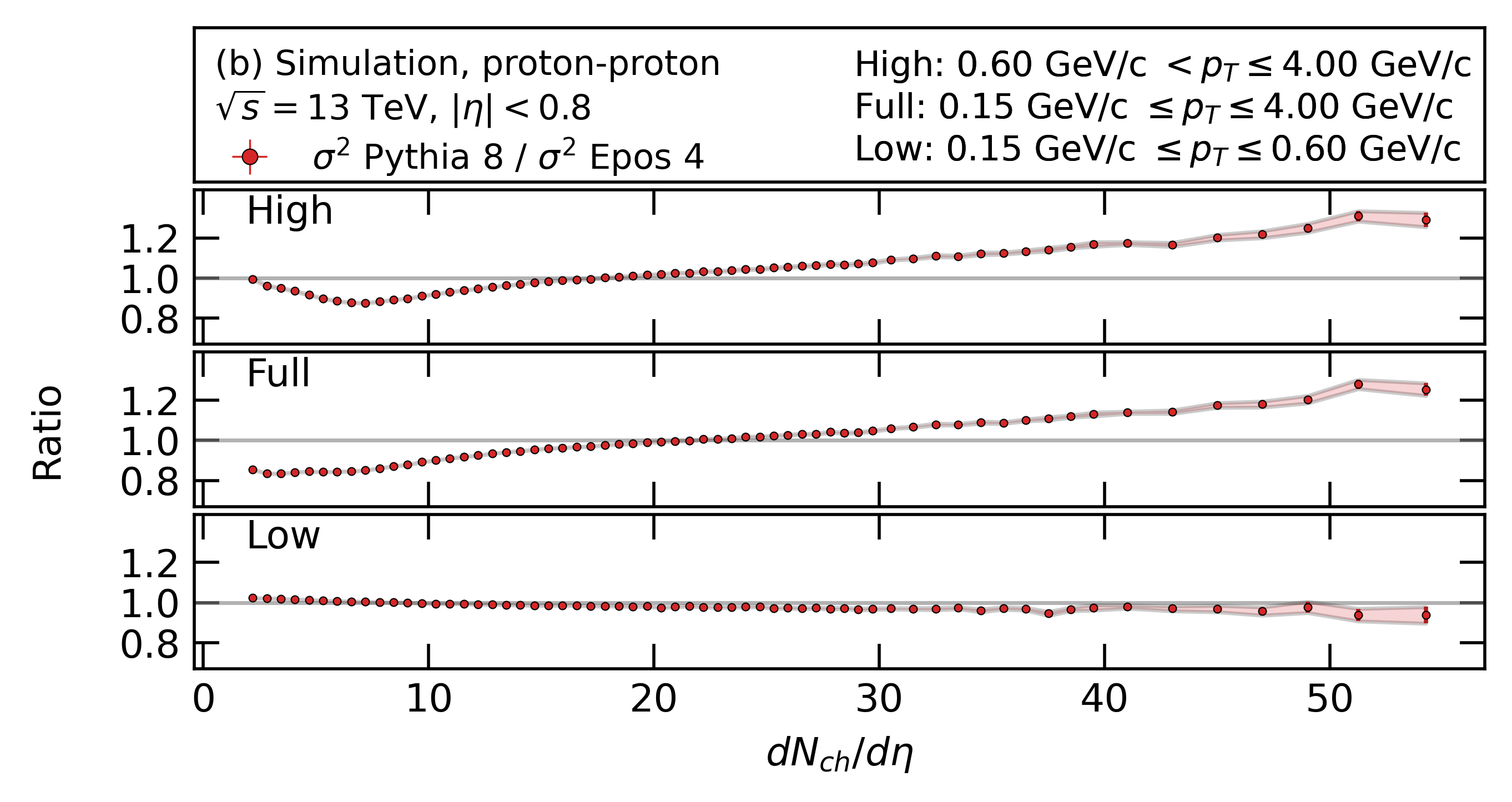}
\caption{
Ratios between {\sc Pythia~8} and {\sc Epos~4} predictions with default settings for (a) \mpT and (b) $\sigma^2$ as a function of \emult~in inelastic \pp collisions at $\sqrt{s}$~=~13 TeV. Ratios are presented for the high-$\pt$ region with $0.6$~GeV/$c$ $< \pt \leq 4$~GeV/$c$ (top panels), full-$\pt$ region with $0.15$~GeV/$c$ $\leq \pt \leq 4.0$~GeV/$c$ (middle panels) and low-$\pt$ region of $0.15$~GeV/$c$ $\leq \pt \leq 0.6$~GeV/$c$ (bottom panels). Statistical uncertainties on the MC predictions are indicated by error bands. 
}
\label{fig:MC_comparisons2}
\end{figure}

We observe a striking feature in Fig.~\ref{fig:pT_means}: both \mpT and variance of the $\pt$ spectra remain constant with respect to \emult~within the low-$\pt$ region. This observation is corroborated by {\sc Pythia~8} simulations. Furthermore, within this $\pt$ range, both \mpT and $\sigma^2$ show no dependence on the centre-of-mass energy, spanning from 0.9 to 13 TeV. To further explore this phenomenon, we compare the predictions of {\sc Pythia~8} to those of {\sc Epos~4} for \mpT and $\sigma^2$ in the three $\pt$ regions at $\sqrt{s}=13$~TeV, as shown in Fig.~\ref{fig:MC_comparisons2}.

\section{Discussion}
\label{sec:discussion}
While reanalysing the experimental charged particles $\pt$ spectra published by the ALICE Collaboration, as presented in Fig.~\ref{fig:pp_normed2}, a consistent crossing-point at around $\sim$0.6~GeV/$c$ transverse momenta value was observed when comparing the ratio of the spectra to the inclusive distribution. This pattern, observed across different \emult~classes and over a wide range of collision energies, is also present in the predictions from both {\sc Pythia~8} and {\sc Epos~4} models, as can be seen in Fig.~\ref{fig:MC_comparisons}.

The separation of the charged particle $\pt$ spectra into two regions defined by the observed crossing-point resulted in two markedly different behaviours of \mpT in each region. The results shown in Fig.~\ref{fig:pT_means} indicate the absence of any variation with \emult~and/or collision energy in the values of \mpT and $\sigma^2$ within the soft, low-$\pt$ region. Our results are consistent with the hypothesis of centre-of-mass energy invariance within the low-$\pt$ part of the spectra, a phenomenon observed by the CDF collaboration in \pp collisions at $\sqrt{s}$~=~630~GeV and 1.8~TeV~\cite{CDF:2001hmt}. On the other hand, the increase of \mpT as a function of multiplicity is moderate for the high-$\pt$ region~\cite{ALICE:2013rdo}.

While based on significantly different underlying assumptions, where {\sc Pythia~8} is rooted in perturbative QCD and string fragmentation and {\sc Epos~4} incorporates collective effects and hydrodynamic considerations, the results in Fig.~\ref{fig:MC_comparisons2} demonstrate that both Monte Carlo models exhibit the same independence of \mpT and $\sigma^2$ on \emult~within the low-$\pt$ range. 

The $\pt$ value around 0.6 GeV/$c$ seems to act as a characteristic momentum scale across different multiplicities. At this momentum, the production rates from various multiplicity classes converge to that of the inclusive one, suggesting a transition in the particle production mechanisms. The region below 0.6 GeV/$c$ is predominantly associated with soft processes, while the region beyond comprises a mixture of soft and hard ones. Therefore, this low-$\pt$ region serves as an excellent laboratory for studying soft physics in \pp collisions. 

We observe that the low-$\pt$ region yields the same {\it mean} values. 
Figure~\ref{fig:pT_means} shows that neither \mpT nor $\sigma^2$ are sensitive to the multiplicity variations in the same region.

Finally, we advocate for the use of statistical parameters of higher order, such as variance and/or skewness~\cite{Giacalone:2020lbm} to identify the differences between different models, Monte Carlo simulations, and experimental data. This is particularly important since the weight of soft and hard processes may differ in the models, and yet resulting in the same mean value. The comparison in Fig.~\ref{fig:pT_means} does indeed point to difference in the behaviors of the two statistical parameters.

\section{Conclusions}
\label{sec:summary}

The observed trends lead us to claim that indeed we are in presence of two different modes withe fact that the soft part of the interactions are reflecting a a mode of interaction common to all collision energies and multiplicities at the LHC. It is important to underline the importance of this conclusion since there are many  attempts to study the collective effects that do extend the range of considered $\pt$ ranges well above the 0.6 GeV/$c$, risking important contributions of collisions of the hard nature 

We have conducted a detailed analysis of the multiplicity dependence of normalised transverse momentum distributions of charged particles, using the extensive dataset of publicly available \pp collision data at $\sqrt{s}$~=~0.9, 5.02, 7 and 13 TeV from the ALICE Collaboration. 

We observe that the mean transverse momentum \mpT is highly sensitive to the selected $\pt$ range, and it is not sensitive within the soft region (low-$p_{\rm T}$) region studied of $0.15$~GeV/$c$ $\leq \pt \leq 0.6$~GeV/$c$. Our investigation reveals a so-far unidentified phenomenon: a universal crossing point of the ratios to the inclusive spectrum at $ \pt \approx 0.6$~GeV/$c$. Below 0.6~GeV/$c$, \mpT exhibits an intriguing invariance from the charged-particle pseudorapidity densities and collision energy. 

This result is supported by {\sc Pythia~8} with different underlying event and generator settings, as well as by {\sc Epos~4}. 
We examine a higher-order statistical parameter, such as variance, to gain further insights into how well the Monte Carlo models agree among themselves and with experimental data.

The absence of any significant \mpT and $\sigma^2$ variation in the low-$\pt$ region 
suggests that the selection of $\pt$ ranges in analyses demands greater scrutiny and justification to avoid potentially misleading conclusions. 
Our study thus contributes to the ongoing controversy regarding the presence of collective effects in small systems. 
Finally, we would like to emphasise that the current conclusions apply to the existing \emult~range measured with sufficient statistics by the ALICE Collaboration. Future LHC runs should enable us to extend our investigation beyond the current multiplicity reach.

\section*{Acknowledgments}
The research was supported by the Hungarian National Research, Development and Innovation Office OTKA K135515, 2019-2.1.11-T\'ET-2019-00078, 2021-4.1.2-NEMZ\_KI-2024-00031, 2021-4.1.2-NEMZ\_KI-2024-00033, 2021-4.1.2-NEMZ\_KI-2022-00008 and 2021-4.1.2-NEMZ\_KI-2022-00009 grants, and by the Wigner Scientific Computing Laboratory and the HUN-REN Wigner Cloud. Support for this work has been received from the Mexican National Council of Humanities, Sciences and Technologies CONAHCYT under Grants No. CF-2042 and No. A1-S-22917. L.S. acknowledges the support received from CONAHCYT for the postdoctoral fellowship. The authors are grateful to Andreas Morsch for his critical comments.

\section*{Data availability}
Data sets generated during the current study are available from the corresponding author on reasonable request.

\bibliography{GP_meanpT_EPJST_2024}

\end{document}